\documentclass[twoside,12pt]{article}
\usepackage{CJK}
\usepackage{indentfirst}
\usepackage{bm}
\usepackage{epsfig}
\usepackage{graphicx}
\usepackage{epstopdf}
\usepackage{color}
\usepackage{mdsymbol}
\newcommand{\upcite}[1]{\textsuperscript{{\cite{#1}}}}
\begin{document}

\begin{CJK*}{GBK}{song}

\thispagestyle{empty} \vspace*{0.8cm}\hbox
to\textwidth{\vbox{\hfill\huge\sf Commun. Theor. Phys.\hfill}}
\par\noindent\rule[3mm]{\textwidth}{0.2pt}\hspace*{-\textwidth}\noindent
\rule[2.5mm]{\textwidth}{0.2pt}


\begin{center}
\LARGE\bf Nonequilibrium effects of reactive flow based on gas kinetic theory$^*$\footnote{}
\end{center}

\footnotetext{\hspace*{-.45cm}\footnotesize $^*$This work is supported by the National Natural Science Foundation of China (NSFC) under Grant No. 51806116.}
\footnotetext{\hspace*{-.45cm}\footnotesize $^\dag$Chuandong Lin, E-mail: linchd3@mail.sysu.edu.cn }

\begin{center}
\rm Xianli Su, and  Chuandong Lin ${^\dagger}$
\end{center}

\begin{center}
\begin{footnotesize} \sl
Sino-French Institute of Nuclear Engineering and Technology, Sun Yat-Sen University, Zhuhai 519082, China.
\end{footnotesize}
\end{center}

\begin{center}
\footnotesize (Received XXXX; revised manuscript received XXXX)

\end{center}

\vspace*{2mm}

\begin{center}
\begin{minipage}{15.5cm}
\parindent 20pt\footnotesize
How to accurately probe chemically reactive flows with essential thermodynamic nonequilibrium effects is an open issue. Via the Chapman-Enskog analysis, the local nonequilibrium particle velocity distribution function is derived from the gas kinetic theory. It is demonstrated theoretically and numerically that the distribution function depends on the physical quantities and derivatives, and is independent of the chemical reactions directly. Based on the simulation results of the discrete Boltzmann model, the departure between equilibrium and nonequilibrium distribution functions is obtained and analyzed around the detonation wave. Besides, it has been verified for the first time that the kinetic moments calculated by summations of the discrete distribution functions are close to those calculated by integrals of their original forms.
\end{minipage}
\end{center}

\begin{center}
\begin{minipage}{15.5cm}
\begin{minipage}[t]{2.3cm}{\bf Keywords:}\end{minipage}
\begin{minipage}[t]{13.1cm}
Discrete Boltzmann method, Reactive flow, Detonation, Nonequilibrium effect.
\end{minipage}\par\vglue8pt

\end{minipage}
\end{center}

\vspace*{5mm}

Chemical reactive flow is a complex physicochemical phenomenon which is ubiquitous in aerospace, energy and power fields, etc.\upcite{oran2005,Law2006}. It exhibits multiscale characteristics in temporal and spatial scales, incorporates various hydrodynamic and thermodynamic nonequilibrium effects\upcite{Nagnibeda2009}. The nonequilibrium effects exert significant influences on fluid systems especially in extremely complex environments\upcite{Nagnibeda2009}, such as the spacecraft reentry into the atmosphere\upcite{regan1993}, multi-component reactive flow in porous media\upcite{juanes2008}, fuel cells\upcite{chen2018,wu2010AP}, phase separation\upcite{Gan2015SF}, hydrodynamic instability\upcite{Lai2016PRE}, and detonation\upcite{Lin2018CNF}. At present, how to accurately probe, predict and analyze chemical reactive flows with essential nonequilibrium effects is still an open issue.

Actually, there are various classes of methodologies to retain the information of velocity distribution functions for fluid systems. For example, on the basis of the distribution function, Nagnibeda et al. established the kinetic theory of transport processes and discussed the features of complex system strongly deviating from the thermal and chemical equilibrium\upcite{Nagnibeda2009}. Besides, on the microscopic level, the distribution function can be obtained by using the direct simulation Monte Carlo\upcite{cercignani1999POF,shoja2021POF}, or molecular dynamics\upcite{zhakhovskii1997JOE,dubey2013PRE}.
As a kinetic mesoscopic methodology, the discrete Boltzmann method (DBM) is a special discretization of the Boltzmann equation in particle velocity space, and has been successfully developed to recover and probe the velocity distribution functions of nonequilibrium physical systems\upcite{Lin2018CNF,Lin2014CTP,lin2014PRE,ZYD2018FOP,LCD2019,lin2020E}.

In fact, the DBM is based on statistical physics and regarded as a variant of the traditional lattice Boltzmann method (LBM)\upcite{Novozhilov2013NHT,LiQing2016,Ashna2017,Yan2019Ele,Sun2020mma}. Compare to standard LBMs, the DBM can address more issues, in particular to simulate the compressible fluid systems with significant nonequilibrium effects\upcite{Lin2014CTP,lin2020E,XuYan2013,Lin2015PRE,Lin2016CNF,Lin2017SR,Lin2018CAF,Gan2018PRE}. At present, there are two means to recover the velocity distribution functions. One relies on the analysis of the detailed nonequilibrium physical quantities to obtain the main features of the velocity distribution function in a qualitative way\upcite{Lin2018CNF,Lin2014CTP,lin2014PRE,LCD2019}. The other is to recover the detailed velocity distribution function by means of macroscopic quantities and their spatio derivatives quantitatively, which can be derived by using the Chapman-Enskog expansion\upcite{ZYD2018FOP,lin2020E}. The two methods are consistent with each other\upcite{lin2020E}.

In the rest of this paper, we firstly derive the nonequilibrium velocity distribution function of reactive fluid based on the Boltzmann equation. Secondly, we give a brief introduction of the DBM for compressible reactive flows. Thirdly, the DBM is utilized to investigate the kinetic moments of the velocity distribution function around the detonation wave. Finally, the nonequilibrium and equilibrium distribution functions as well as their differences are obtained and analyzed.

Now, let us introduce the popular Bhatanger-Gross-Krook (BGK) Boltzmann equation,
\begin{equation}
\frac{\partial f}{\partial t}\text{+}\mathbf{v}\cdot \nabla f=\frac{1}{\tau }\left( f-{{f}^{eq}} \right)+R
\tt{,}
\label{BGKmodel}
\end{equation}
where ${\tau}$ denotes the relaxation time, $t$ the time, $f$ the velocity distribution function.
The equilibrium distribution function\upcite{LCD2019} is
\begin{equation}
{{f}^{eq}}=n{{\left( \frac{1}{2\pi T} \right)}^{D/2}}{{\left( \frac{1}{2\pi IT} \right)}^{1/2}}\exp \left[ -\frac{{{\left| \mathbf{v}-\mathbf{u} \right|}^{2}}}{2T}-\frac{{{\eta }^{2}}}{2IT} \right]
\label{feq}
\tt{,}
\end{equation}
where $D=2$ denotes the dimensional translational degree of freedom, $I$ stands for extra degrees of freedom due to vibration and/or rotation, and $\eta$ represents the corresponding vibrational and/or rotational energies. Here $n$ is the particle number density, $\mathbf{u}$ the hydrodynamic velocity, $T$ the temperature, $m=1$ the particle mass, and $\rho = n m$ the mass density. 

On the right-hand side of Eq. (\ref{BGKmodel}), $R$ is the chemical term describing the change rate of the distribution function due to chemical reactions, i.e.,
\begin{equation}
	R = {{\left. \frac{d f}{d t} \right|}_{R}}
	\label{ReactionTerm1}
	\tt{.}
\end{equation}
To derive the explicit expression of the chemical term, the following qualifications are assumed\upcite{Lin2015PRE}: $t_{mr} < t_{cr} < t_{sys}$, where $t_{mr}$, $t_{cr}$ and $t_{sys}$ represent the time scale of molecular relaxation, the time scale of chemical reaction and the characteristic time scale of the system, respectively. Under the condition $t_{mr} < t_{cr}$, Eq. (\ref{ReactionTerm1}) can be approximated by
\begin{equation}
	R\approx {{\left. \frac{d{{f}^{eq}}}{dt} \right|}_{R}}=\frac{\partial {{f}^{eq}}}{\partial \rho }{{\left. \frac{\partial \rho }{\partial t} \right|}_{R}}+\frac{\partial {{f}^{eq}}}{\partial \mathbf{u}}{{\left. \frac{\partial \mathbf{u}}{\partial t} \right|}_{R}}+\frac{\partial {{f}^{eq}}}{\partial T}{{\left. \frac{\partial T}{\partial t} \right|}_{R}}
	\label{ReactionTerm2}
	\tt{,}
\end{equation}
as $f^{eq}$ is the function of $\rho$, $\mathbf{u}$, and $T$, respectively. 
Furthermore, the assumption $t_{cr} < t_{sys}$ leads to the following conclusion: the chemical reaction results in the change of temperature directly, as the density and flow velocity remain unchanged during the rapid reaction process. Consequently, Eq. (\ref{ReactionTerm2}) can be reduced to
\begin{equation}
	R=\frac{\partial {{f}^{eq}}}{\partial T}{{\left. \frac{\partial T}{\partial t} \right|}_{R}}
	\label{ReactionTerm3}
	\tt{.}
\end{equation}
Substituting Eq. (\ref{feq}) into Eq. (\ref{ReactionTerm3}) gives
\begin{equation}
R={{f}^{eq}}\times \left[ -\frac{D+1}{2T}+\frac{{{\left| \mathbf{v}-\mathbf{u} \right|}^{2}}}{2{{T}^{2}}}+\frac{{{\eta }^{2}}}{2I{{T}^{2}}} \right]\frac{2Q{\lambda }'}{D+I}
\label{ReactionTerm_Expression}
\tt{,}
\end{equation}
where $Q$ indicates the chemical heat release per unit mass of fuel, ${\lambda }'$ is the change rate of the mass fraction of chemical product. Additionally, a two-step reaction scheme is employed to mimic the essential dynamics of a chain-branching reaction of detonation in this paper\upcite{Ng2005CTM}. 

Via the Chapman-Enskog analysis, we derive the first-order approximation formula of the velocity distribution function of reacting flows through the macroscopic quantities and their spatial and temporal derivatives,
\begin{align}
& f={{f}^{eq}}-{{\tau }}\left[ \frac{\partial {{f}^{eq}}}{\partial \rho }\left( \frac{\partial \rho }{\partial {{t}}}+{{v}_{\alpha }}\frac{\partial \rho }{\partial {{r}_{\alpha }}} \right) \right. \nonumber \\
& \left. +\frac{\partial {{f}^{eq}}}{\partial T}\left( \frac{\partial T}{\partial {{t}}}+{{v}_{\alpha }}\frac{\partial T}{\partial {{r}_{\alpha }}} \right)+\frac{\partial {{f}^{eq}}}{\partial {{u}_{\beta }}}\left( \frac{\partial {{u}_{\beta }}}{\partial {{t}}}+{{v}_{\alpha }}\frac{\partial {{u}_{\beta }}}{\partial {{r}_{\alpha }}} \right) \right] \nonumber \\
& +{{\tau }}\left( \frac{\partial {{f}^{eq}}}{\partial T}{{\left. \frac{\partial T}{\partial {{t}}} \right|}_{R}} \right)
\label{velocity distribution function-1}
\tt{,}
\end{align}
in terms of
\begin{equation}
\frac{\partial {{f}^{eq}}}{\partial \rho }={{f}^{eq}}\times \frac{1}{\rho }
\label{feq-rho}
\tt{,}
\end{equation}
\begin{equation}
\frac{\partial {{f}^{eq}}}{\partial T}={{f}^{eq}}\times \left[ -\frac{D+1}{2T}+\frac{{{\left|\mathbf{v}-\mathbf{u} \right|}^{2}}}{2{{T}^{2}}}+\frac{{{\eta }^{2}}}{2I{{T}^{2}}} \right]
\label{feq-T}
\tt{,}
\end{equation}
and
\begin{equation}
\frac{\partial {{f}^{eq}}}{\partial {{u}_{\beta }}}={{f}^{eq}}\times \frac{({{v}_{\beta }}-{{u}_{\beta }})}{T}
\label{feq-u}
\tt{.}
\end{equation}
Note that the change rate of temperature consists of two parts, i.e.,
\begin{equation}
\frac{\partial T}{\partial {{t}}}\text{=}{{\left. \frac{\partial T}{\partial {{t}}} \right|}_{R}}-\frac{2T}{D+I}\frac{\partial {{u}_{\alpha }}}{\partial {{r}_{\alpha }}}-{{u}_{\alpha }}\frac{\partial T}{\partial {{r}_{\alpha }}}
\label{T-t}
\tt{,}
\end{equation}
on the right-hand side of which the first term describes the part caused by the heat release of chemical reactions,
\begin{equation}
{{\left. \frac{\partial T}{\partial {{t}}} \right|}_{R}}\text{=}\frac{2Q{\lambda }'}{D+I}
\label{T-tr}
\tt{,}
\end{equation}
and the other two terms reflect the parts due to the spatial gradients of velocity and temperature.

Therefore, Eq. (\ref{velocity distribution function-1}) can be simplified as
\begin{align}
	& f={{f}^{eq}}-{{\tau }}\left[ \frac{\partial {{f}^{eq}}}{\partial \rho }\left( \frac{\partial \rho }{\partial {{t}}}+{{v}_{\alpha }}\frac{\partial \rho }{\partial {{r}_{\alpha }}} \right) \right. \nonumber \\
	& \text{+}\frac{\partial {{f}^{eq}}}{\partial T}\left( -\frac{2T}{D+I}\frac{\partial {{u}_{\alpha }}}{\partial {{r}_{\alpha }}}-{{u}_{\alpha }}\frac{\partial T}{\partial {{r}_{\alpha }}}+{{v}_{\alpha }}\frac{\partial T}{\partial {{r}_{\alpha }}} \right) \nonumber \\
	& \left. \text{+}\frac{\partial {{f}^{eq}}}{\partial {{u}_{\beta }}}\left( \frac{\partial {{u}_{\beta }}}{\partial {{t}}}+{{v}_{\alpha }}\frac{\partial {{u}_{\beta }}}{\partial {{r}_{\alpha }}} \right) \right]
	\label{velocity distribution function-2}
	\tt{,}
\end{align}
with
\begin{equation}
     \frac{\partial \rho }{\partial t}=-\rho \frac{\partial {{u}_{\alpha }}}{\partial {{r}_{\alpha }}}-{{u}_{\alpha }}\frac{\partial \rho }{\partial {{r}_{\alpha }}} 
     \tt{.}
     \label{density partial}
\end{equation}

It can be found from Eqs. (\ref{T-t}), (\ref{T-tr}) and (\ref{density partial}) that the temporal derivatives can be expressed by the spatial derivatives. Those formulas are obtained from the Chapman-Enskog expansion. 
In fact, there are two ways to calculate the simulation results of $\frac{\partial \rho }{\partial t}$ or $\frac{\partial T }{\partial t}$. 
One method is to calculate the temporal change rate directly. For example,
\[\frac{\partial \rho }{\partial t}=\frac{{{\rho }^{t+\Delta t}}-{{\rho }^{t-\Delta t}}}{2\Delta t} \tt{.}\]
The other method is to use Eqs. (\ref{T-t}) or (\ref{density partial})
where the spatial derivatives can be computed by the finite difference scheme. The results given by the two methods are similar to each other.

Moreover, it can be inferred from Eq. (\ref{velocity distribution function-2}) that the chemical reaction term is eliminated, so it has no contribution to the velocity distribution function directly. This is due to the aforementioned assumption that the chemical time scale is longer than the molecular relaxation time. The case where the chemical time scale is close to or less than the molecular relaxation time is not considered in this work.

\begin{figure}[tbp]
\begin{center}
	\includegraphics[bbllx=0pt,bblly=0pt,bburx=145pt,bbury=146pt,width=0.3\textwidth]{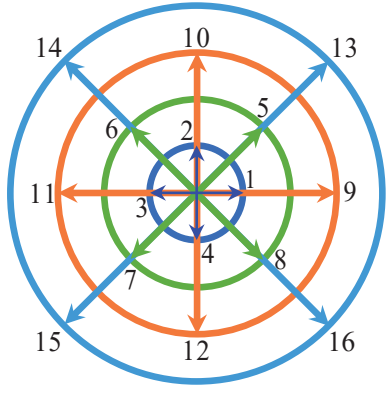}
\end{center}
\caption{Sketch for the discrete velocity model.}
\label{Fig01}
\end{figure}

In this work, the BGK DBM is used to mimic and measure the nonequilibrium reactive flows\upcite{LCD2019}.
The discretization of the model in particle velocity space takes the form
\begin{equation}
\frac{\partial {{f}_{i}}}{\partial t}+{{v}_{i\alpha }}\frac{\partial {{f}_{i}}}{\partial {{r}_{\alpha }}}=\frac{1}{\tau }\left( {{f}_{i}}-f_{i}^{eq} \right)+{{R}_{i}}
\tt{,}
\label{DiscreteBoltzmannEquation}
\end{equation}
where $f_i$ and $f_i^{eq}$ represent the discrete distribution function and its equilibrium counterpart, respectively. $v_i$ denotes the discrete velocity with $i=1$, $2$, $3$, $\cdots$, $N$, and $N = 16$ is the total number of discrete velocities. Here a two-dimensional sixteen-velocity model is employed, see Fig. \ref{Fig01}.

Physically, the DBM is approximately equivalent to a continuous fluid model plus a coarse-grained model for discrete effects. Meanwhile, the DBM is roughly equivalent to a hydrodynamic model plus a coarse-grained model of thermodynamic nonequilibrium behaviors. For the sake of recovering the NS equations in the hydrodynamic limit, the discrete equilibrium distribution functions
$f^{eq}_i$ are required to satisfy the following relationship,
\begin{equation}
\iint{{{f}^{eq}}\Psi d\mathbf{v}d\eta }=\sum\nolimits_{i}{f_{i}^{eq}}\Psi_{i}
\label{MomentFeq}
\tt{,}
\end{equation}
with the particle velocities $\Psi = 1$, $\mathbf{v}$, $\left( \mathbf{v}\cdot \mathbf{v}+{{\eta }^{2}} \right)$, $\mathbf{vv}$, $\left( \mathbf{v}\cdot \mathbf{v}+{{\eta }^{2}} \right)\mathbf{v}$, $\mathbf{vvv}$, $\left( \mathbf{v}\cdot \mathbf{v}+{{\eta }^{2}} \right)\mathbf{vv}$,
and the corresponding discrete velocities,
$\Psi_{i} = 1$, ${\mathbf{v}}_{i}$, $\left( {{\mathbf{v}}_{i}}\cdot {{\mathbf{v}}_{i}}+\eta _{i}^{2} \right)$, ${{\mathbf{v}}_{i}}{{\mathbf{v}}_{i}}$, $\left( \mathbf{v}_{i}\cdot \mathbf{v}_{i}+\eta _{i}^{2} \right){{\mathbf{v}}_{i}}$, ${{\mathbf{v}}_{i}}{{\mathbf{v}}_{i}}{{\mathbf{v}}_{i}}$, $\left( {{\mathbf{v}}_{i}}\cdot {{\mathbf{v}}_{i}}+\eta _{i}^{2} \right){{\mathbf{v}}_{i}}{{\mathbf{v}}_{i}}$.

Furthermore, one merit of the DBM is to capture nonequilibrium information described by the following (but not limited to) high-order kinetic moments
\begin{equation}
{{\mathbf{M}}_{2}}=\sum\limits_{i}{{{f}_{i}}{{\mathbf{v}}_{i}}}{{\mathbf{v}}_{i}}
\label{M_2}
\tt{,}
\end{equation}
\begin{equation}
{{\mathbf{M}}^{eq}_{2}}=\sum\limits_{i}{{{f}^{eq}_{i}}{{\mathbf{v}}_{i}}}{{\mathbf{v}}_{i}}
\label{M^eq_2}
\tt{,}
\end{equation}
\begin{equation}
{{\mathbf{M}}_{3,1}}=\sum\limits_{i}{{{f}_{i}}\left( {{\mathbf{v}}_{i}}\cdot {{\mathbf{v}}_{i}}+{{\eta }^2_{i}} \right){{\mathbf{v}}_{i}}}
\label{M_3,1}
\tt{,}
\end{equation}
\begin{equation}
{{\mathbf{M}}^{eq}_{3,1}}=\sum\limits_{i}{{{f}^{eq}_{i}}\left( {{\mathbf{v}}_{i}}\cdot {{\mathbf{v}}_{i}}+{{\eta }^2_{i}} \right){{\mathbf{v}}_{i}}}
\label{M^eq_3,1}
\tt{,}
\end{equation}
\begin{equation}
	{{\mathbf{M}}_{3}}=\sum\limits_{i}{{{f}_{i}}{{\mathbf{v}}_{i}}}{{\mathbf{v}}_{i}}{{\mathbf{v}}_{i}}
	\label{M_3}
	\tt{,}
\end{equation}
\begin{equation}
\mathbf{M}_{3}^{eq}=\sum\limits_{i}{f_{i}^{eq}{{\mathbf{v}}_{i}}}{{\mathbf{v}}_{i}}{{\mathbf{v}}_{i}}
	\label{M^eq_3}
	\tt{,}
\end{equation}
\begin{equation}
{{\mathbf{M}}_{4,2}}=\sum\limits_{i}{{{f}_{i}}\left( {{\mathbf{v}}_{i}}\cdot {{\mathbf{v}}_{i}}+\eta _{i}^{2} \right){{\mathbf{v}}_{i}}}{{\mathbf{v}}_{i}}
	\label{M_4,2}
	\tt{,}
\end{equation}
\begin{equation}
	\mathbf{M}_{4,2}^{eq}=\sum\limits_{i}{f_{i}^{eq}\left( {{\mathbf{v}}_{i}}\cdot {{\mathbf{v}}_{i}}+\eta _{i}^{2} \right){{\mathbf{v}}_{i}}}{{\mathbf{v}}_{i}}
	\label{M^eq_4,2}
	\tt{,}
\end{equation}
\begin{equation}
{{\mathbf{\Delta }}_{2}} = {{\mathbf{M}}_{2}}-\mathbf{M}_{2}^{eq} 
\label{delta_2}
\tt{,}
\end{equation}
\begin{equation}
{{\mathbf{\Delta }}_{3,1}} = {{\mathbf{M}}_{3,1}}-\mathbf{M}_{3,1}^{eq}
\label{delta_3,1}
\tt{,}
\end{equation}
\begin{equation}
{{\mathbf{\Delta }}_{3}}={{\mathbf{M}}_{3}}-\mathbf{M}_{3}^{eq}
	\label{delta_3}
	\tt{,}
\end{equation}
\begin{equation}
{{\mathbf{\Delta }}_{4,2}}={{\mathbf{M}}_{4,2}}-\mathbf{M}_{4,2}^{eq}
	\label{delta_4,2}
	\tt{,}
\end{equation}
where the ${\mathbf{M}}_{2}$, ${{\mathbf{M}}_{3,1}}$, ${\mathbf{M}}_{3}$ and ${\mathbf{M}}_{4,2}$ are the kinetic moments of the distribution functions,  ${\mathbf{M}}^{eq}_{2}$, ${{\mathbf{M}}^{eq}_{3,1}}$, ${{\mathbf{M}}^{eq}_{3}}$ and ${{\mathbf{M}}^{eq}_{4,2}}$ denote the corresponding equilibrium counterparts,  ${{\mathbf{\Delta }}_{2}}$, ${{\mathbf{\Delta }}_{3,1}}$, ${{\mathbf{\Delta }}_{3}}$ and ${{\mathbf{\Delta }}_{4,2}}$ are the differences between them. Here, ${{\mathbf{\Delta }}_{2}}$ represents the viscous stress tensor and nonorganized momentum flux, ${{\mathbf{\Delta }}_{3,1}}$ and ${{\mathbf{\Delta }}_{3}}$ are relevant to the nonorganized energy fluxes. ${{\mathbf{\Delta }}_{4,2}}$ is related to the flux of nonorganized energy flux.

To verify the consistency of theoretical and numerical results of the nonequilibrium manifestations of reactive flows, firstly, we simulate a reaction process in a uniform resting system. 
The specific-heat ratio is $\gamma =5/3$, the chemical heat release $Q=1$, the space step $\Delta x=\Delta y=5\times {{10}^{-5}}$, the time step $\Delta t=2\times {{10}^{-6}}$, and the discrete velocities ($v_a$, $v_b$, $v_c$, $v_d$, $\eta_a$, $\eta_b$, $\eta_c$, $\eta_d$) = ($3.7$, $3.2$, $1.4$, $1.4$, $2.4$, $0$, $0$, $0$), respectively. In order to possess a high computational efficiency, only one mesh grid (${{N}_{x}}\times {{N}_{y}}=1\times1$) is used, and the periodic boundary condition is adopted in each direction, because the physical field is uniformly distributed. It is found that all simulated nonequilibrium physical quantities (including ${{\mathbf{\Delta }}_{2}}$, ${{\mathbf{\Delta }}_{3,1}}$, ${{\mathbf{\Delta }}_{3}}$, and ${{\mathbf{\Delta }}_{4,2}}$) remain zero during the evolution. Therefore, in the process of the chemical reaction, the deviation of velocity distribution function $f$ from its equilibrium counterpart $f^{eq}$ is zero, i.e $f=f^{eq}$. 
Besides, all physical gradients are zero in the simulation process due to the uniform distribution of physical quantities. Consequently, it is numerically verified that the chemical reaction does not contribute to the nonequilibrium effects directly.$^*$\footnote{} 
\footnotetext{\hspace*{-.45cm}\footnotesize $^*$It should be mentioned that the chemical reaction may change the physical gradients which make an impact on the nonequilibrium effects. In other words, the chemical reaction plays an indirect role in nonequilibrium effect of the reactive flows.} 
This result is consistent with the aforementioned theory that the distribution function depends on the physical quantities and derivatives, and is independent of chemical reactions directly, see Eq. (\ref{velocity distribution function-2}).

For the purpose of further validation, the one-dimensional ($1$-D) steady detonation is simulated. The initial configuration, obtained from the Hugoniot relation, takes the form
\begin{equation}
\left\{ \begin{array}{*{35}{l}}
	{{\left(\rho, {u}_{x}, {u}_{y}, T, \xi, \lambda \right)}_{L}} = \left( 1.38549, 0.70711, 0, 2.01882, 1, 1 \right) \tt{,} \\
	{{\left(\rho, {u}_{x}, {u}_{y}, T, \xi, \lambda \right)}_{R}} = \left( 1, 0, 0, 1, 0, 0 \right)
	\nonumber
	\tt{,}
\end{array} \right.
\end{equation}
where the subscript $L$ indicates $0\le x\le 0.00555$, and $R$ indicates $0.00555<x\le 0.555$. The Mach number is $1.96$.
To ensure the resolution is high enough, the grid is chosen as ${{N}_{x}}\times {{N}_{y}}=11100\times 1$, other parameters are the same as before.
Furthermore, the inflow and/or outflow boundary conditions are employed in the $x$ direction, and the periodic boundary condition is adopted in the $y$ direction.
\begin{figure}[tbp]
\begin{center}
	\includegraphics[bbllx=10pt,bblly=5pt,bburx=560pt,bbury=462pt,width=0.68\textwidth]{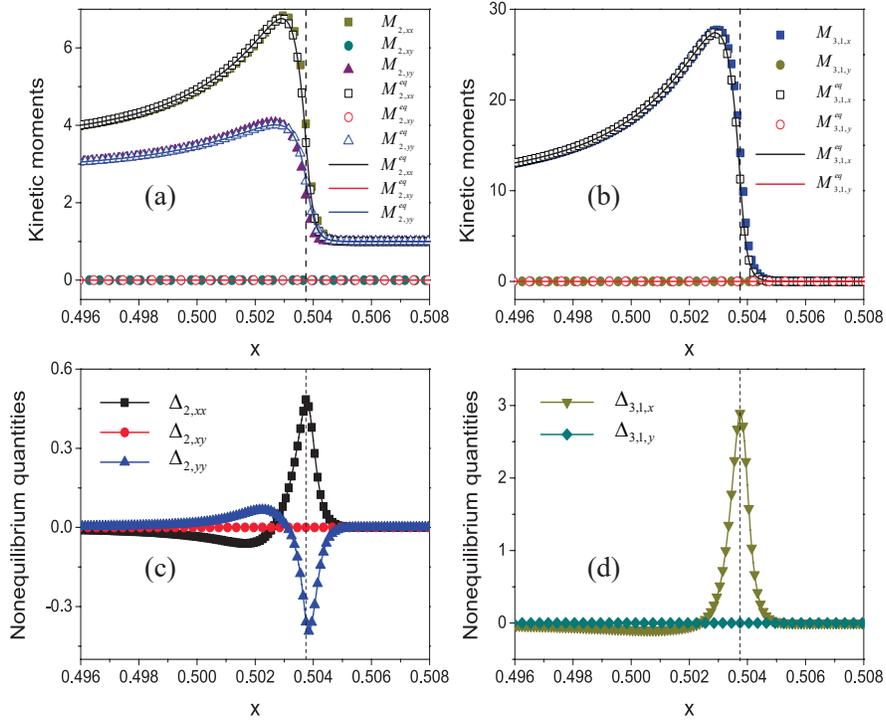}
\end{center}
\caption{The nonequilibrium and equilibrium kinetic moments, and the differences between them. (a)-(d) shows the independent variables of ${\mathbf{M}_{2}}$, ${\mathbf{M}_{3,1}}$, ${{\mathbf{\Delta }}_{2}}$, and ${{\mathbf{\Delta }}_{3,1}}$, respectively. The legends are in each plot.}
\label{Fig02}
\end{figure}

Figure \ref{Fig02} displays the kinetic moments of velocity distribution function (${M_{2,xx}}$, ${M_{2,xy}}$, ${M_{2,yy}}$, ${M_{3,1,x}}$, ${M_{3,1,y}}$), the equilibrium counterparts (${M^{eq}_{2,xx}}$, ${M^{eq}_{2,xy}}$, ${M^{eq}_{2,yy}}$, ${M^{eq}_{3,1,x}}$, ${M^{eq}_{3,1,y}}$), and the nonequilibrium quantities (${{\Delta }_{2,xx}}$, ${{\Delta }_{2,xy}}$, ${{\Delta }_{2,yy}}$, ${{\Delta }_{3,1,x}}$,$ {{\Delta }_{3,1,y}}$) around the detonation front. Here ${{\Delta }_{2,xx}}$ represents twice the nonorganized energy in the $x$ degree of freedom, and ${{\Delta }_{2,yy}}$ twice the nonorganized energy in the $y$ degree of freedom. ${{\Delta }_{3,1,x}}$ and ${{\Delta }_{3,1,y}}$ denote twice the nonorganized energy fluxes in the $x$ and $y$ directions, respectively. The dashed line is located at the position $x=0.50375$.

In Fig. \ref{Fig02} (a), the solid squares, circles and triangles stand for the DBM results of kinetic moments of distribution function ${M_{2,xx}}, {M_{2,xy}}$, and ${M_{2,yy}}$, respectively. The hollow squares, circles and triangles represent the DBM results of the kinetic moments of equilibrium distribution function ${M^{eq}_{2,xx}}$, ${M^{eq}_{2,xy}}$, and ${M^{eq}_{2,yy}}$, respectively. And the solid lines indicate the corresponding analytic solutions, ${M^{eq}_{2,xx}}=\rho \left( T+u_{x}^{2} \right)$, ${M^{eq}_{2,xy}}=\rho {{u}_{x}}{{u}_{y}}$, and ${M^{eq}_{2,yy}}=\rho \left( T+u_{y}^{2} \right)$, respectively.
With the detonation wave propagating from left to right, ${M_{2,xx}}$, ${M^{eq}_{2,xx}}$, ${M_{2,yy}}$, ${M^{eq}_{2,yy}}$ first increase due to the compressible effect, then decrease owing to the rarefaction effect, and form a peak around the detonation front. Meanwhile, ${M_{2,xy}}$ and ${M^{eq}_{2,xy}}$ remain zero, because the detonation wave propagates forwards in the $x$ direction. In addition, as for the equilibrium kinetic moments (${M^{eq}_{2,xx}}$, ${M^{eq}_{2,xy}}$, and ${M^{eq}_{2,yy}}$), the DBM results are in good agreement with the analytical solutions.

In Fig. \ref{Fig02} (b), the solid squares and circles represent ${M_{3,1,x}}$ and ${M_{3,1,y}}$, respectively. The hallow symbols denote the equilibrium counterparts ${M^{eq}_{3,1,x}}$, and ${M^{eq}_{3,1,y}}$. And the solid lines indicate the analytic solutions ${M^{eq}_{3,1,x}}=\rho {{u}_{x}}\left[ \left( D+I+2 \right)T+{{u}^{2}} \right]$ and ${M^{eq}_{3,1,y}}=\rho {{u}_{y}}\left[ \left( D+I+2 \right)T+{{u}^{2}} \right]$.
Similarly, ${M_{3,1,x}}$ and ${M^{eq}_{3,1,x}}$ ascend rapidly and then decline slowly due to the compressible and rarefaction effects, respectively. ${M_{3,1,y}}$ and ${M^{eq}_{3,1,y}}$ are still zero in the one-dimensional simulation.

In Fig. \ref{Fig02} (c), as the detonation wave travels from left to right, ${{\Delta }_{2,xx}}$ expressed by the solid line with squares first increases, then decreases, and increases afterwards, so it forms a high positive peak and a negative trough. Actually, Fig. \ref{Fig02} (c) is consistent with Fig \ref{Fig02} (a) where ${M_{2,xx}}$ first greater than ${M^{eq}_{2,xx}}$ and then less than ${M^{eq}_{2,xx}}$ as the detonation wave travels forwards. Physically, ${{\Delta }_{2,xx}}$ stands for twice the nonorganized energy in the $x$ direction, its positive peak and the negative trough correspond to the compression and rarefaction effects, respectively. The solid line with triangles denotes twice the nonorganized energy in the $y$ direction ${{\Delta }_{2,yy}}$, which first decreases to form a negative trough and then increases to form a low positive peak. This trend is also consistent with the results of ${M_{2,yy}}$ and ${M^{eq}_{2,yy}}$ in the Fig. \ref{Fig02} (a). Additionally, ${{\Delta }_{2,xy}} ={M_{2,xy}} - {M^{eq}_{2,xy}} = 0$ in Fig. \ref{Fig02} (c) is in line with ${M_{2,xy}} = {M^{eq}_{2,xy}} = 0$ in Fig. \ref{Fig02} (a).

In Fig. \ref{Fig02} (d), ${{\Delta }_{3,1,x}}$ and ${{\Delta }_{3,1,y}}$ denote twice the nonorganized energy fluxes in the $x$ and $y$ directions, respectively. ${{\Delta }_{3,1,x}}$ and ${{\Delta }_{3,1,y}}$ are the differences between the kinetic moment ${M_{3,1,x}}, {M_{3,1,y}}$ and their equilibrium counterparts ${M^{eq}_{3,1,x}}, {M^{eq}_{3,1,y}}$, respectively. Obviously, ${{\Delta }_{3,1,x}}$ forms a positive peak and then a low negative trough. Because the ${M_{3,1,x}}$ first greater and then less than ${M^{eq}_{3,1,x}}$. Besides, the ${M_{3,1,y}}$ and ${M^{eq}_{3,1,y}}$ are zero, which causes ${{\Delta }_{3,1,y}} = {M_{3,1,y}} - {M^{eq}_{3,1,y}}$ to be zero as well.

Next, let us verify that the kinetic moments calculated by the summations of the discrete distribution functions are close to those calculated by integrals of their original forms at the location $x=0.50375$. The kinetic moments calculated by the summations of the discrete distribution functions are (${{\Delta }_{2,xx}}$, ${{\Delta }_{2,xy}}$, ${{\Delta }_{2,yy}}$, ${{\Delta }_{3,1,x}}$, ${{\Delta }_{3,1,y}}$) = ($0.48449$, $0$, $-0.35844$, $2.89123$, $0$), while the results of the corresponding integration counterparts are (${{\Delta }_{2,xx}}$, ${{\Delta }_{2,xy}}$, ${{\Delta }_{2,yy}}$, ${{\Delta }_{3,1,x}}$, ${{\Delta }_{3,1,y}}$) = ($0.56488$, $0$, $-0.27255$, $2.80181$, $0$). The relative errors are ($17\%$, $0\%$, $24\%$, $3\%$, $0\%$), which is roughly satisfactory. For the first time, this test demonstrates the accuracy of the nonequilibrium manifestations measured by the DBM, and validates the consistence of the DBM with its theoretical basis.

\begin{figure}[tbp]
\begin{center}
	\includegraphics[bbllx=0pt,bblly=0pt,bburx=560pt,bbury=591pt,width=0.66\textwidth]{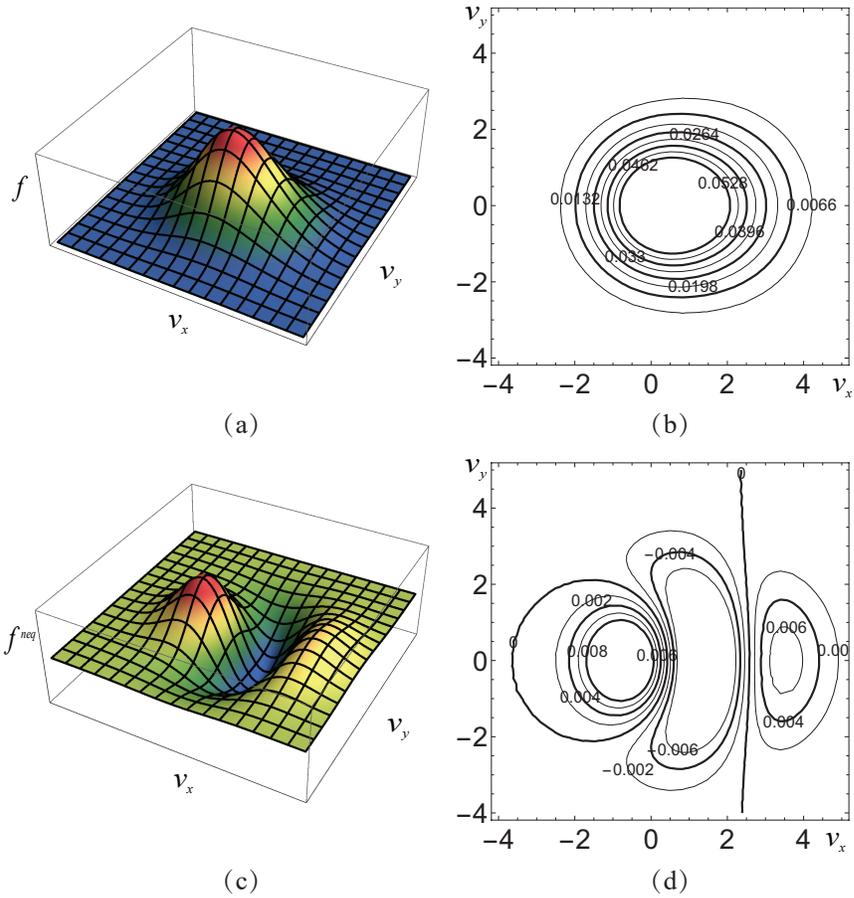}
\end{center}
\caption{The velocity distribution function (a) and its corresponding contour(b), the deviation of the velocity distribution function from the equilibrium state (c) and its corresponding contour (d).}
\label{Fig03}
\end{figure}

To further perform a quantitative study of the nonequilibrium state around the detonation wave, Fig. \ref{Fig03} (a) displays the velocity distribution function at the peak of ${{\Delta }_{2,xx}}$, which is on the vertical dashed line in Fig. \ref{Fig02}. It is clear that the velocity distribution function has a peak in the two-dimensional velocity space. Actually, due to the nonequilibrium effects, the velocity distribution function deviates from its local equilibrium counterpart, i.e., the Maxwellian velocity distribution function.

In order to have an intuitive study of the local velocity distribution function, Fig. \ref{Fig03} (b) shows its contours in the velocity space, which is in line with Fig. \ref{Fig03} (a). Clearly, the peak is asymmetric in the $v_x$ direction and symmetric in the $v_y$ direction. The contour lines are close to each other near the peak (especially on the left side), and becomes sparse away from the peak (especially on the right side). That is to say, the gradient is sharp near the peak (on the left side in especial), and smooth far from the peak (on the left side in especial).

To have a deep understanding of the deviation of the velocity distribution function from the equilibrium state, Fig. \ref{Fig03} (c) depicts the difference between the nonequilibrium and equilibrium distribution functions in the two-dimensional velocity space. It is obvious that there are both positive and negative deviations around the detonation wave. Along the $v_x$ direction, a high positive peak first appears, then decreases to form a valley, and then increases to a low positive peak.

As can be seen in Fig. \ref{Fig03} (d), the deviation is symmetric about $v_{y} = 0$, and asymmetric about $v_{x} = u_{x}$.
The contour plot consists of three segments along the $v_x$ direction. The leftmost segment is in the region of the first peak, where the contour lines are approximately elliptical. The middle part is in the low valley area that seems like a ``moon" shape. And the rightmost one is in the low peak area, which likes a cobblestone. The contour lines between the high peak and the valley are closer to each other than those between the valley and low peak, because the gradients between the leftmost and middle parts are sharp than those between the middle and rightmost regions.

\begin{figure}[tbp]
\begin{center}
	\includegraphics[bbllx=1pt,bblly=0pt,bburx=569pt,bbury=420pt,width=0.66\textwidth]{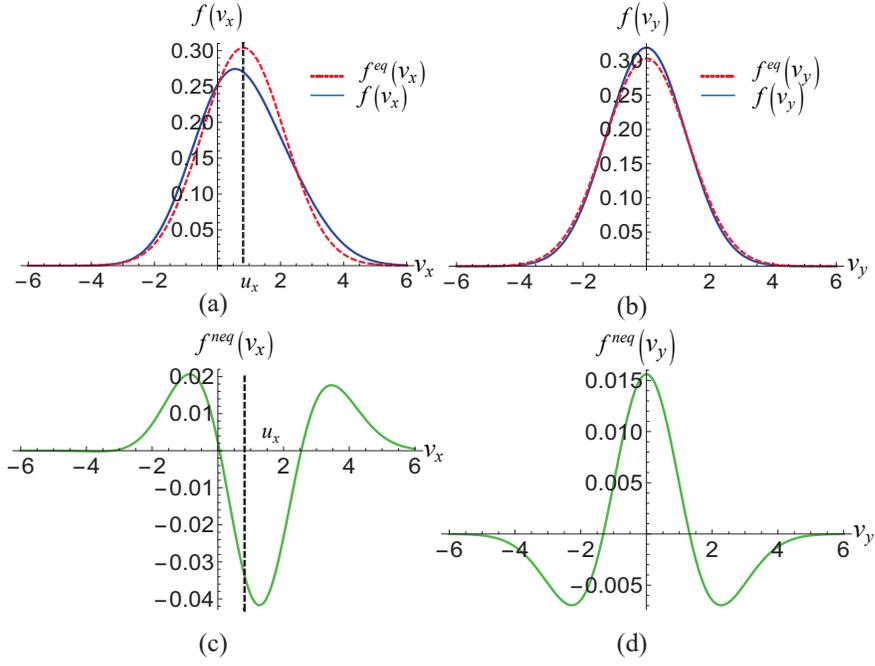}
\end{center}
\caption{One-dimensional nonequilibrium and equilibrium distribution functions in the ${v_x}$ (a) and ${v_y}$ (b) directions, and the differences between them in the ${v_x}$ (c) and ${v_y}$ (d) directions.}
\label{Fig04}
\end{figure}

Finally, let us investigate the one-dimensional distribution functions and the corresponding deviations from the equilibrium states. Figures \ref{Fig04} (a) and (b) depict the velocity distribution functions in the $v_x$ and $v_y$ directions, respectively. The solid lines represent the velocity distribution functions $f({{v}_{x}})=\iint{fd}{{v}_{y}}d\eta $ and $f({{v}_{y}})=\iint{fd}{{v}_{x}}d\eta $, the dashed curves express the equilibrium counterparts ${{f}^{eq}}({{v}_{x}})=\iint{{{f}^{eq}}d}{{v}_{y}}d\eta $ and ${{f}^{eq}}({{v}_{y}})=\iint{{{f}^{eq}}d}{{v}_{x}}d\eta $, respectively. Figures \ref{Fig04} (c) and (d) show ${f^{neq}}({{v}_{x}}) = f({{v}_{x}}) - {{f}^{eq}}({{v}_{x}})$ and ${f^{neq}}({{v}_{y}}) = f({{v}_{y}}) - {{f}^{eq}}({{v}_{y}})$ which indicate the departures of distribution functions from the equilibrium state in the $v_x$ and $v_y$ directions, respectively. The following points can be obtained.

(I) In Figs. \ref{Fig04} (a) - (b), there is a peak for each curve of $f({{v}_{x}})$, ${{f}^{eq}}({{v}_{x}})$, $f({{v}_{y}})$, and ${{f}^{eq}}({{v}_{y}})$. In Figs. \ref{Fig04} (c) - (d), there are two peaks and a trough for ${f^{neq}}({{v}_{x}})$, while a peak and two troughs for ${f^{neq}}({{v}_{y}})$. Along the $v_x$ direction, ${f^{neq}}({{v}_{x}})$ forms a positive peak firstly, then decreases to form a valley, and then increases to a second positive peak. Because ${{f}}({{v}_{x}})$ is first greater than ${{f}^{eq}}({{v}_{x}})$, then less than ${{f}^{eq}}({{v}_{x}})$, and finally greater than ${{f}^{eq}}({{v}_{x}})$ again. Similarly, the relation $f({{v}_{y}}) > {{f}^{eq}}({{v}_{y}})$ or $f({{v}_{y}}) < {{f}^{eq}}({{v}_{y}})$ in Fig. \ref{Fig04} (b) leads to the results ${f^{neq}}({{v}_{y}}) > 0$ or ${f^{neq}}({{v}_{y}}) < 0$ in Fig. \ref{Fig04} (d).

(II)  ${f}({{v}_{x}})$ and ${f^{neq}}({{v}_{x}})$ are asymmetric about the vertical dashed line located at $v_x = {u_x}$, while ${{f}^{eq}}({{v}_{x}})$ is symmetric. Physically, as the detonation evolves, the compressible effect plays a significant role in the front of the detonation wave, and the internal energy in the $x$ degree of freedom increases faster than in other degrees of freedom, and there exists nonorganized heat flux in the $x$ direction.

(III) In Figs. \ref{Fig04} (b) and (d), each curve of $f({{v}_{y}})$, ${{f}^{eq}}({{v}_{y}})$ and ${f^{neq}}({{v}_{y}})$ has a positive peak which is symmetric about ${v_y = 0}$. On the left and right parts of ${f^{neq}}({{v}_{y}})$ are two identical troughs that are symmetrically distributed in Fig. \ref{Fig04} (b). Because the periodic boundary condition is imposed on the $y$ direction, the equilibrium and nonequilibrium velocity distributions for $v_{y} > 0$ and $v_{y} < 0$ are symmetrical.

(IV) The nonequilibrium manifestations in Figs. \ref{Fig02} (a)-(d) are consistent with the deviations of distribution functions in Figs. \ref{Fig04} (a)-(d). Specifically, the trend of ${f^{neq}}({{v}_{x}})$ indicates that ${{f}}({{v}_{x}})$ is ``fatter" and ``lower" than ${{f}^{eq}}({{v}_{x}})$, which means the nonorganized momentum flux ${{\Delta }_{2,xx}}>0$. The trend of ${f^{neq}}({{v}_{y}})$ means that ${f}({{v}_{y}})$ is ``thinner" and ``higher" than ${f^{eq}}({{v}_{y}})$, which indicates ${{\Delta }_{2,yy}}<0$. Meanwhile, the portion ${{f}}({{v}_{x}} > {{u}_{x}})$ is ``fatter" than the part ${{f}}({{v}_{x}} < {{u}_{x}})$, which is named ``positive skewness" and indicates ${{\Delta }_{3,1,x}}>0$. And the symmetry of ${f^{neq}}({{v}_{y}})$ means ${{\Delta }_{3,1,y}}=0$.

In conclusion, via the Chapman-Enskog expansion, the velocity distribution function of compressible reactive flows is expressed by using the macroscopic quantities and their spatial derivatives. The equilibrium and nonequilibrium distribution functions in one- and two-dimensional velocity spaces are recovered quantitatively from the physical quantities of the DBM, which is an accurate and efficient gas kinetic method. The departure between the equilibrium and nonequilibrium distribution functions is in line with the nonequilibrium quantities measured by the DBM. Moreover, it is for the first time to verify that the kinetic moments measured by summations of the distribution function resemble those assessed by integrals of the original forms, which consists with the theoretical basis of the DBM. 
In addition, under the condition that the chemical time scale is longer than the molecular relaxation time,
it is numerically and theoretically demonstrated that the chemical reaction imposes no direct impact on the thermodynamic nonequilibrium effects.

\newpage
\vspace*{-1mm}
\begin{small}\baselineskip=10pt\itemsep-2pt
\bibliography{References}	

\begin{thebibliography}{10}

\bibitem{oran2005}
Elaine~S Oran and Jay~P Boris.
\newblock {\em Numerical simulation of reactive flow}.
\newblock Cambridge university press, 2005.

\bibitem{Law2006}
C.~K. Law.
\newblock {\em Combustion physics}.
\newblock Cambridge University Press, Cambridge, 2006.

\bibitem{Nagnibeda2009}
E.~Nagnibeda and E.~Kustova.
\newblock {\em Non-equilibrium reacting gas flows: kinetic theory of transport
  and relaxation processes}.
\newblock Springer, Berlin, 2009.

\bibitem{regan1993}
Frank~J Regan.
\newblock {\em Dynamics of atmospheric re-entry}.
\newblock Aiaa, 1993.

\bibitem{juanes2008}
Ruben Juanes.
\newblock Nonequilibrium effects in models of three-phase flow in porous media.
\newblock {\em Advances in Water Resources}, 31(4):661--673, 2008.

\bibitem{chen2018}
Li~Chen, Mengyi Wang, Qinjun Kang, and Wenquan Tao.
\newblock Pore scale study of multiphase multicomponent reactive transport
  during co2 dissolution trapping.
\newblock {\em Advances in water resources}, 116:208--218, 2018.

\bibitem{wu2010AP}
Hao Wu, Peter Berg, and Xianguo Li.
\newblock Steady and unsteady 3d non-isothermal modeling of pem fuel cells with
  the effect of non-equilibrium phase transfer.
\newblock {\em Applied Energy}, 87(9):2778--2784, 2010.

\bibitem{Gan2015SF}
Yanbiao Gan, Aiguo Xu, Guangcai Zhang, and Sauro Succi.
\newblock Discrete boltzmann modeling of multiphase flows: Hydrodynamic and
  thermodynamic non-equilibrium effects.
\newblock {\em Soft Matter}, 11(26):5336--5345, 2015.

\bibitem{Lai2016PRE}
Huilin Lai, Aiguo Xu, Guangcai Zhang, Yanbiao Gan, Yangjun Ying, and Sauro
  Succi.
\newblock Nonequilibrium thermohydrodynamic effects on the rayleigh-taylor
  instability in compressible flows.
\newblock {\em Physical Review E}, 94(2):023106, 2016.

\bibitem{Lin2018CNF}
Chuandong Lin and Kai~H Luo.
\newblock Mesoscopic simulation of nonequilibrium detonation with discrete
  boltzmann method.
\newblock {\em Combustion and Flame}, 198:356--362, 2018.

\bibitem{cercignani1999POF}
Carlo Cercignani, Aldo Frezzotti, and Patrick Grosfils.
\newblock The structure of an infinitely strong shock wave.
\newblock {\em Physics of fluids}, 11(9):2757--2764, 1999.

\bibitem{shoja2021POF}
Ahmad Shoja-Sani, Ehsan Roohi, and Stefan Stefanov.
\newblock Homogeneous relaxation and shock wave problems: Assessment of the
  simplified and generalized bernoulli trial collision schemes.
\newblock {\em Physics of Fluids}, 33(3):032004, 2021.

\bibitem{zhakhovskii1997JOE}
VV~Zhakhovskii, K~Nishihara, and SI~Anisimov.
\newblock Shock wave structure in dense gases.
\newblock {\em Journal of Experimental and Theoretical Physics Letters},
  66(2):99--105, 1997.

\bibitem{dubey2013PRE}
Awadhesh~Kumar Dubey, Anna Bodrova, Sanjay Puri, and Nikolai Brilliantov.
\newblock Velocity distribution function and effective restitution coefficient
  for a granular gas of viscoelastic particles.
\newblock {\em Physical Review E}, 87(6):062202, 2013.

\bibitem{Lin2014CTP}
Chuandong Lin, Aiguo Xu, Guangcai Zhang, and Yingjun Li.
\newblock Polar coordinate lattice boltzmann kinetic modeling of detonation
  phenomena.
\newblock {\em Communications in Theoretical Physics}, 62(5):737, 2014.

\bibitem{lin2014PRE}
Chuandong Lin, Aiguo Xu, Guangcai Zhang, Yingjun Li, and Sauro Succi.
\newblock Polar-coordinate lattice boltzmann modeling of compressible flows.
\newblock {\em Physical Review E}, 89(1):013307, 2014.

\bibitem{ZYD2018FOP}
YuDong Zhang, Ai-Guo Xu, Guang-Cai Zhang, Zhi-Hua Chen, and Pei Wang.
\newblock Discrete ellipsoidal statistical bgk model and burnett equations.
\newblock {\em Frontiers of Physics}, 13(3):1--13, 2018.

\bibitem{LCD2019}
Chuandong Lin and Kai~H Luo.
\newblock Discrete boltzmann modeling of unsteady reactive flows with
  nonequilibrium effects.
\newblock {\em Physical Review E}, 99(1):012142, 2019.

\bibitem{lin2020E}
Chuandong Lin, Xianli Su, and Yudong Zhang.
\newblock Hydrodynamic and thermodynamic nonequilibrium effects around shock
  waves: Based on a discrete boltzmann method.
\newblock {\em Entropy}, 22(12):1397, 2020.

\bibitem{Novozhilov2013NHT}
Vasily Novozhilov and Conor Byrne.
\newblock Lattice boltzmann modeling of thermal explosion in natural convection
  conditions.
\newblock {\em Numerical Heat Transfer, Part A: Applications}, 63(11):824--839,
  2013.

\bibitem{LiQing2016}
Qing Li, Kai~Hong Luo, QJ~Kang, YL~He, Q~Chen, and Q~Liu.
\newblock Lattice boltzmann methods for multiphase flow and phase-change heat
  transfer.
\newblock {\em Progress in Energy and Combustion Science}, 52:62--105, 2016.

\bibitem{Ashna2017}
Mostafa Ashna, Mohammad~Hassan Rahimian, and Abbas Fakhari.
\newblock Extended lattice boltzmann scheme for droplet combustion.
\newblock {\em Physical Review E}, 95(5):053301, 2017.

\bibitem{Yan2019Ele}
WW~Yan, YF~Yuan, JY~Xiang, Y~Wu, TY~Zhang, SM~Yin, and SY~Guo.
\newblock Construction of triple-layered sandwich nanotubes of carbon@
  mesoporous tio2 nanocrystalline@ carbon as high-performance anode materials
  for lithium-ion batteries.
\newblock {\em Electrochimica Acta}, 312:119--127, 2019.

\bibitem{Sun2020mma}
Rui Du, Jincheng Wang, and Dongke Sun.
\newblock Lattice-boltzmann simulations of the convection-diffusion equation
  with different reactive boundary conditions.
\newblock {\em Mathematics}, 8(1):13, 2020.

\bibitem{XuYan2013}
Bo~Yan, Aiguo Xu, Guangcai Zhang, Yangjun Ying, and Hua Li.
\newblock Lattice boltzmann model for combustion and detonation.
\newblock {\em Frontiers of Physics}, 8(1):94--110, 2013.

\bibitem{Lin2015PRE}
Aiguo Xu, Chuandong Lin, Guangcai Zhang, and Yingjun Li.
\newblock Multiple-relaxation-time lattice boltzmann kinetic model for
  combustion.
\newblock {\em Physical Review E}, 91(4):043306, 2015.

\bibitem{Lin2016CNF}
Chuandong Lin, Aiguo Xu, Guangcai Zhang, and Yingjun Li.
\newblock Double-distribution-function discrete boltzmann model for combustion.
\newblock {\em Combustion and Flame}, 164:137--151, 2016.

\bibitem{Lin2017SR}
Chuandong Lin, Kai~Hong Luo, Linlin Fei, and Sauro Succi.
\newblock A multi-component discrete boltzmann model for nonequilibrium
  reactive flows.
\newblock {\em Scientific reports}, 7(1):14580, 2017.

\bibitem{Lin2018CAF}
Chuandong Lin and Kai~Hong Luo.
\newblock Mrt discrete boltzmann method for compressible exothermic reactive
  flows.
\newblock {\em Computers \& Fluids}, 166:176--183, 2018.

\bibitem{Gan2018PRE}
Yanbiao Gan, Aiguo Xu, Guangcai Zhang, Yudong Zhang, and Sauro Succi.
\newblock Discrete boltzmann trans-scale modeling of high-speed compressible
  flows.
\newblock {\em Physical Review E}, 97(5):053312, 2018.

\bibitem{Ng2005CTM}
H.~D. Ng, M.~I. Radulescu, A.~J. Higgins, N.~Nikiforakis, and J.~H.~S. Lee.
\newblock Numerical investigation of the instability for one-dimensional
  chapman-jouguet detonations with chain-branching kinetics.
\newblock {\em Combustion Theory and Modelling}, 9(3):385--401, 2005.

\end{thebibliography}

\end{small}
\end{CJK*}
\end{document}